\documentclass[runningheads]{llncs}
\usepackage{fancybox}
\usepackage{hyperref}
\usepackage{amsmath,amsfonts,latexsym}
\usepackage{amstext}
\usepackage{amssymb}
\usepackage{bm}
\usepackage{mathrsfs}
\usepackage[all]{xy}
\xyoption{v2}

\usepackage{graphicx}
\usepackage{color}

\newcommand\establish[1]{\ensuremath{{\vartriangleleft}{\blacktriangleright}}}

\newcommand\elementgen[1]{\ensuremath{#1}}

\newcommand\elementc{\elementgen{c}}

\newcommand\elementi{\elementgen{i}}

\newcommand\setgen[1]{\ensuremath{#1}}

\newcommand\setC{\setgen{C}}

\newcommand\setI{\setgen{I}}

\newcommand\mapgen[1]{\ensuremath{#1}}

\newcommand\mapf{\mapgen{f}}

\newcommand\vertexgen[1]{\ensuremath{#1}}

\newcommand\vertexv{\vertexgen{v}}

\newcommand\edgegen[1]{\ensuremath{#1}}
\newcommand\edgee{\edgegen{e}}

\newcommand\graphgen[1]{\ensuremath{#1}}
\newcommand\graphA{\graphgen{A}}

\newcommand\graphG{\graphgen{G}}

\newcommand\graphI{\graphgen{I}}

\newcommand\graphK{\graphgen{K}}

\newcommand\graphTG{\graphgen{TG}}

\newcommand\nodesgen[1]{\ensuremath{\graphgen{#1}_V}}
\newcommand\edgesgen[1]{\ensuremath{\graphgen{#1}_E}}
\newcommand\sourcegen[1]{\ensuremath{sc^{\graphgen{#1}}}}
\newcommand\targetgen[1]{\ensuremath{tg^{\graphgen{#1}}}}
\newcommand\nodesA{\nodesgen{A}}

\newcommand\nodesG{\nodesgen{G}}

\newcommand\edgesA{\edgesgen{A}}

\newcommand\edgesG{\edgesgen{G}}

\newcommand\sourceA{\sourcegen{A}}

\newcommand\targetA{\targetgen{A}}

\newcommand\graphlonggen[1]{\ensuremath{(\nodesgen{#1},\edgesgen{#1},\sourcegen{#1},\targetgen{#1})}}

\newcommand\graphlongG{\graphlonggen{G}}

\newcommand\katgen[1]{\ensuremath{\mathsf{#1}}}

\newcommand\katC{\katgen{C}}

\newcommand\katgraph{\katgen{Graph}}

\newcommand\katset{\katgen{Set}}

\newcommand\objectsgen[1]{\ensuremath{\katgen{#1}_V}}

\newcommand\identityof[1]{\ensuremath{id_{\objectgen{#1}}}}

\newcommand\morphismgen[1]{\ensuremath{#1}}

\newcommand\morphismf{\morphismgen{f}}

\newcommand\objectgen[1]{\ensuremath{#1}}

\newcommand\objectC{\objectgen{C}}

\newcommand\functorcat[2]{\ensuremath{[#1\to #2]}}

\newcommand\functorgen[1]{\ensuremath{\mathtt{#1}}}

\newcommand\calK{\ensuremath{\mathcal{K}}}

\newcommand{\arity}[2][2]{\ensuremath{\alpha^{\sig{#1}[#2]}}}

\renewcommand\katgen[1]{\ensuremath{\mathbf{#1}}}

\newcommand{\setpred}{\setgen{\Pi}}
\newcommand{\katvar}{\katgen{Var}}
\newcommand{\katcontext}{\katgen{Cxt}}

\newcommand{\katsketch}{\katgen{Sk}}
\newcommand{\katmsketch}{\katgen{mSk}}

\renewcommand\objectsgen[1]{\ensuremath{\katgen{#1}_{Obj}}}

\newcommand\objectstype{\objectsgen{Var}}

\newcommand{\metasig}{\ensuremath{\Xi}}

\newcommand\featuregen[1]{\ensuremath{\mathtt{#1}}}

\newcommand\featurecomp{\featuregen{cmp}}
\newcommand\featureequ{\featuregen{equ}}
\newcommand\featureid{\featuregen{id}}
\newcommand\featuremonic{\featuregen{mon}}

\newcommand\featurepb{\featuregen{pb}}
\newcommand\featurefinal{\featuregen{fnl}}
\newcommand\featurepo{\featuregen{po}}
\newcommand\featureprod{\featuregen{prod}}

\newcommand{\predP}{\ensuremath{P}}

\newcommand{\condition}{\ensuremath{c}}
\newcommand{\conditiona}{\ensuremath{c_1}}
\newcommand{\conditionb}{\ensuremath{c_2}}

\newcommand{\conditionct}{\ensuremath{ct}}
\newcommand{\conditiong}{\ensuremath{g}}

\newcommand\contextof{\ensuremath{\blacktriangleright}}

\newcommand\propimplication[2]{\ensuremath{(#1\rightarrow #2)}}

\newcommand\setofsketchcond[1]{\ensuremath{\functorSC(#1)}}

\newcommand\true{\ensuremath{\bm{T}}}
\newcommand\false{\ensuremath{\bm{F}}}
\newcommand\guardedexistential[4]{\ensuremath{(#1\rightarrow\bm{\exists}(#2, #3: #4))}}
\newcommand\guardeduniversal[4]{\ensuremath{(#1\rightarrow\bm{\forall}(#2, #3 : #4))}}
\newcommand\uncondexists[3]{\ensuremath{\exists(#1,#2:#3)}}
\newcommand\uncondforall[3]{\ensuremath{\forall(#1,#2:#3)}}

\newcommand\morphismsatisfies[3]{\ensuremath{#1\; \models^{#2} #3}}
\newcommand\morphismsatisfiesnot[3]{\ensuremath{#1\; \not\models^{#2} #3}}
\newcommand\sketchsatisfies[3]{\ensuremath{#1\; \models (#2, #3)}}
\newcommand\sketchsatisfiesnot[3]{\ensuremath{#1\; \not\models (#2, #3)}}

\newcommand{\initialobject}{\ensuremath{\bm{0}}}
\newcommand\initialmorphism[1]{\ensuremath{!_{#1}}}

\newcommand\contextA{\ensuremath{A}}
\newcommand\contextB{\ensuremath{B}}
\newcommand\contextC{\ensuremath{C}}
\newcommand\contextD{\ensuremath{D}}
\newcommand\contextG{\ensuremath{G}}
\newcommand\contextH{\ensuremath{H}}
\newcommand\contextK{\ensuremath{K}}
\newcommand\contextL{\ensuremath{L}}
\newcommand\contextM{\ensuremath{M}}
\newcommand\contextR{\ensuremath{R}}

\newcommand\statement{\ensuremath{st}}

\newcommand\bindinga{\ensuremath{\beta}}

\newcommand\functorCstr{\functorgen{Cstr}}

\newcommand\functorSC{\functorgen{SC}}

\newcommand\functorStm{\functorgen{Stm}}

\newcommand{\contextmorphisma}{\ensuremath{\varphi}}

\newcommand{\contextmorphismc}{\ensuremath{\gamma}}

\newcommand{\contextmorphismm}{\ensuremath{\mu}}
\newcommand{\contextmorphismr}{\ensuremath{\varrho}}
\newcommand{\contextmorphismt}{\ensuremath{\tau}}

\newcommand{\sketchA}{\ensuremath{\mathfrak{A}}}
\newcommand{\sketchB}{\ensuremath{\mathfrak{B}}}
\newcommand{\sketchC}{\ensuremath{\mathfrak{C}}}
\newcommand{\sketchD}{\ensuremath{\mathfrak{D}}}
\newcommand{\sketchG}{\ensuremath{\mathfrak{G}}}
\newcommand{\sketchH}{\ensuremath{\mathfrak{H}}}
\newcommand{\sketchK}{\ensuremath{\mathfrak{K}}}
\newcommand{\sketchL}{\ensuremath{\mathfrak{L}}}
\newcommand{\sketchR}{\ensuremath{\mathfrak{R}}}

\newcommand{\setofconstraintsG}{\ensuremath{C^{\mathfrak{G}}}}

\newcommand\setofstatementsgen[1]{\ensuremath{S^{\mathfrak{#1}}}}
\newcommand{\setofstatementsA}{\setofstatementsgen{A}}
\newcommand{\setofstatementsB}{\ensuremath{S^{\mathfrak{B}}}}

\newcommand{\setofstatementsD}{\ensuremath{S^{\mathfrak{D}}}}
\newcommand{\setofstatementsG}{\ensuremath{S^{\mathfrak{G}}}}

\newcommand{\setofstatementsK}{\ensuremath{S^{\mathfrak{K}}}}
\newcommand{\setofstatementsL}{\ensuremath{S^{\mathfrak{L}}}}
\newcommand{\setofstatementsR}{\ensuremath{S^{\mathfrak{R}}}}

\renewcommand\arity{\ensuremath{\alpha}}

\newcommand\flar[3]{\ensuremath{#1\!:#2\rightarrow #3}}

\newcommand\dlar[3]{\ensuremath{#1:\xymatrix{{#2}\ar@{-|}[r]&{#3}}}} 

\newcommand{\isdef}{:=}

\begin{document}
\pagestyle{plain} 
\title{First-Order Sketch Conditions and Constraints\linebreak -- A Category Independent Approach --}
\author{Uwe Wolter\orcidID{0000-0002-7553-9858}}
\institute{University of Bergen, Norway, \email{Uwe.Wolter@uib.no}}

\maketitle
 
\begin{abstract}
Generalizing different variants of "graph conditions and constraints" 
as well as "universal constraints" and "negative universal constraints" in the Diagram Predicate Framework (DPF) 
, we introduce for arbitrary categories \katcontext\ and "statement" functors \flar{\functorStm}{\katcontext}{\katset} general first-order sketch conditions and constraints.
Sketches are used in DPF to formalize different kinds of 
software models. We
discuss the use of sketch constraints for describing the syntactic structure of sketches. 
We outline the use of sketch constraints to deduce knowledge implicitly given in a sketch as well as procedures to deduce sketch constraints from given sketch constraints. 
We use the simple but paradigmatic 
modeling formalism "Category Theory" as running example.
\end{abstract}

\begin{keywords} Graph conditions $\cdot$ Graph constraints  $\cdot$ Generalized sketches $\cdot$  Sketch conditions $\cdot$  Sketch constraints $\cdot$ Diagram Predicate Framework 
\end{keywords}

\section{Introduction}

The present paper constitutes together with \cite{Wolter2021logics,WolterDK18} the first stage of expansion of a bigger project.
The overall objective of our project is to extend the Diagram Predicate Framework (DPF) \cite{DW08,RRLW09_JLAP,Rutle10,RRLW10_JLAP_NWPT2009,WolterMH2020} by a diagrammatic logic enabling us (1) to describe and reason about the syntactic structure of software models, (2) to deduce knowledge implicitly given in a model, (3) to control and reason about model transformations and (4) to formalize and handle meta-modeling issues. Such a diagrammatic logic should also provide a basis to develop tools for diagrammatic software engineering as well as for diagrammatic reasoning and diagrammatic proofs (supplementing traditional "string based" tools). 

To achieve this goal we have been choosing a more unconventional approach. We neither wanted to encode traditional first-order logic of binary predicates by graphs \cite{Rensink04} nor to emulate nested  graph conditions by traditional first-order formulas \cite{HabelP09}. We developed rather a method to define, in a conservative way, logics of first-order conditions and constraints in arbitrary categories. By "conservative" we mean that the application of our 
universal 
method to different categories of graphs, as in  \cite{BrugginkCHK11,EEPT06,HabelP09,Kosiol0TZ20,Rensink04} for example, allows us to define the various corresponding variants of (nested) graph conditions and constraints. To validate the use of the term "first-order" we have to ensure, in addition, that the application of our method to the category \katset\ results in a logic comprising the essential features of traditional first-order logic.

Sketches are used in DPF to formalize different kinds of diagrammatic software models. In our applications \cite{RLGRW14_JFAC,RRLW09_JLAP,Rutle10,RRLW10_JLAP_NWPT2009}, a sketch is given by an underlying (typed) graph and additional statements of various kinds, called atomic constraints, linked to different parts of the graph. Atomic constraints have to be satisfied by any semantic interpretation of a sketch. In \cite{Wolter2021logics} we have shown how to define arbitray first-order constraints  in arbitrary categories instead of just atomic constraints as in \cite{DW08}. While \cite{Wolter2021logics} addresses, especially, the semantics of sketches defined in arbitrary categories, we focus in this paper on the syntactic structure of sketches. So, in the same way as (nested) graph conditions and constraints have been developed to describe properties of graph morphisms or graphs, respectively, we define first-order sketch conditions and constraints to describe properties of sketch morphisms or sketches, respectively. 

The paper is organized as follows. In Section \ref{sec:abstract-sketches} we construct for arbitrary categories \katcontext\ of contexts and "statement" functors  \flar{\functorStm}{\katcontext}{\katset} a corresponding category of sketches and discuss adhesiveness. Section 3 defines general first-order sketch conditions and their satisfaction by  context morphisms. These conditions give us as well local as global first-order sketch constraints at hand. Section 4 presents vital observations, insights, concepts and ideas for the future development of a deduction calculus for sketch constraints. We show that our universal method enables us to establish and to work with conceptual hierarchies of sketches analogously to \cite{Mak97}. A short conclusion is given in Section \ref{sec:conclusions}.

\section{Abstract sketches}
\label{sec:abstract-sketches}
To give a category independent definition of first-order sketch conditions and constraints, we rely on the syntactic constituents of "institutions of constraints" as introduced in \cite{Wolter2021logics}:
First, we assume a category \katcontext. The objects in \katcontext\ are called  \textbf{contexts} while we refer to the morphisms in \katcontext\ as \textbf{context morphisms}. 
Second, we assume a functor \flar{\functorStm}{\katcontext}{\katset} assigning to each context \contextK\ a set \(\functorStm(\contextK)\) of \textbf{statements in \contextK} and to each context morphisms \flar{\contextmorphisma}{\contextK}{\contextG} a map \flar{\functorStm(\contextmorphisma)}{\functorStm(\contextK)}{\functorStm(\contextG)}. For any statement  \(\statement\in\functorStm(\contextK)\) we will denote the image \(\functorStm(\contextmorphisma)(\statement)\in\functorStm(\contextG)\) also simply by $\contextmorphisma(\statement)$.

An \textbf{(abstract) sketch} $\sketchK=(\contextK,\setofstatementsK)$ is given by a context $\contextK$ and a set $\setofstatementsK\subseteq\functorStm(\contextK)$ of statements in \contextK.
A \textbf{(strict) morphism} $\contextmorphisma:\sketchK\to\sketchG$ between two sketches $\sketchK=(\contextK,\setofstatementsK)$ and $\sketchG=(\contextG,\setofstatementsG)$ is given by a morphism $\contextmorphisma:\contextK\to\contextG$ in \katcontext\ such that $\setofstatementsG\subseteq\contextmorphisma(\setofstatementsK)$. We denote by $\katsketch$ the category of all sketches and all sketch morphisms.  Generalizing the constructions and results in \cite{WolterM2013} we can prove that $\katsketch$ has pushouts and pullbacks as long as \katcontext\ does.
\begin{proposition}
\label{prop:pushouts}
Let $\xymatrix{\sketchB & \sketchC\ar[l]_{\contextmorphismm}\ar[r]^{\contextmorphismr}& \sketchA}$
be a span of sketch morphisms. 
\\
\begin{minipage}[c]{.3\linewidth}
\vspace*{-1ex}
	$$\xymatrix{
		\sketchC \ar[r]^{\contextmorphismr}\ar[d]_(.5){\contextmorphismm}\ar@{}[dr]|{PO}
		& \sketchA  \ar[d]^(.45){\contextmorphismm^*}
		\\
		\sketchB\ar[r]^{\contextmorphismr^*} 
		& \sketchD
	}$$
\end{minipage}
\begin{minipage}[c]{.7\linewidth}
If there exists a pushout $\xymatrix{\contextB\ar[r]^{\contextmorphismr^*} & \contextD& \contextA\ar[l]_(.4){\contextmorphismm^*}}$ of the span 
$\xymatrix{\contextB & \contextC\ar[l]_{\contextmorphismm}\ar[r]^{\contextmorphismr}& \contextA}$ of morphisms in \katcontext, then the diagram, on the left, is a \textbf{pushout} in \katsketch\ where
\begin{equation}\label{eq:pushout-sketches}
\sketchD\isdef(\contextD,\contextmorphismm^*(\setofstatementsA)\cup\contextmorphismr^*(\setofstatementsB))
\end{equation} 
\end{minipage}
\end{proposition}
\begin{proposition}
\label{prop:pullbacks}
Let $\xymatrix{\sketchB\ar[r]^{\contextmorphismm} & \sketchC& \sketchA\ar[l]_(.4){\contextmorphismr}}$
be a cospan of sketch morphisms. 

\noindent
\begin{minipage}[c]{.25\linewidth}
\vspace*{-1ex}
$$\xymatrix{
\sketchD \ar[r]^{\contextmorphismm^*}\ar[d]_(.5){\contextmorphismr^*}\ar@{}[dr]|{PB}
& \sketchA  \ar[d]^(.45){\contextmorphismr}
\\
\sketchB\ar[r]^{\contextmorphismm} 
& \sketchC
}$$
\end{minipage}
\begin{minipage}[c]{.75\linewidth}
If there exists a pullback $\xymatrix{\contextB& \contextD\ar[l]_(.4){\contextmorphismr^*}\ar[r]^(.5){\contextmorphismm^*} & \contextA}$ of the cospan 
$\xymatrix{\contextB\ar[r]^{\contextmorphismm} & \contextC& \contextA\ar[l]_{\contextmorphismr}}$ of morphisms in \katcontext, then the diagram, on the left, is a \textbf{pullback} in \katsketch\ where
\begin{equation}\label{eq:pullback-sketches}
\sketchD\isdef(\contextD,\{\statement\in\functorStm(\contextD)\mid \contextmorphismm^*(\statement)\in\setofstatementsA,\contextmorphismr^*(\statement)\in\setofstatementsB \} )
\end{equation} \end{minipage}
\end{proposition}

\begin{remark}[Adhesiveness]
\label{rem:adhesiveness}
The category \katsketch\ will be, in general, not adhesive, even if \katcontext\ is adhesive, since \setofstatementsD\  in Proposition \ref{prop:pushouts} is not constructed by a pushout in \katset\ and in Proposition \ref{prop:pullbacks} not by a pullback in \katset\ either. 

To repair this deficiency, we can work with "multi sketches" where statements do have their own identity. A \textbf{multi sketch} \(\sketchK=(\contextK,I^\sketchK,stm^\sketchK)\) is given by a context \contextK, a set \(I^\sketchK\) of identifiers and a map \(stm^\sketchK:I^\sketchK\to\functorStm(\contextK)\). A \textbf{morphism} $(\contextmorphisma,\mapf):\sketchK\to\sketchG$ between two multi sketches $\sketchK$ and $\sketchG$ is given by a morphism $\contextmorphisma:\contextK\to\contextG$ in \katcontext\ and a map \(\mapf:I^\sketchK\to I^\sketchG\) such that $\contextmorphisma(stm^\sketchK(i))=stm^\sketchG(\mapf(i))$ for all \(i\in I^\sketchK\). Pushouts in the category \katmsketch\ of multi sketches can be always constructed by componentwise pushouts of contexts in \katcontext\ and of sets of identifiers in \katset, respectively. To ensure that componentwise pullbacks in \katcontext\ and \katset, respectively, give us a pullback in \katmsketch, we have to assume, however, that the functor \flar{\functorStm}{\katcontext}{\katset} preserves pullbacks. This is the case for any "logic of first-order constraints" \cite{Wolter2021logics} and thus for any variant of contexts and statements we used and may use, in the future, in DPF \cite{RLGRW14_JFAC,RRLW09_JLAP,Rutle10,RRLW10_JLAP_NWPT2009} (see Example \ref{ex:dpf-sketches}). 

If \functorStm\ preserves pullbacks, the monomorphisms in \katmsketch\ are exactly the componentwise monomorphisms and \katmsketch\ becomes adhesive if \katcontext\ is adhesive.
\qed
\end{remark}

\begin{example}[DPF: Sketches]
\label{ex:dpf-sketches}
In DPF we can define a variant of sketches for any 
category \katcontext, any subcategory $\katvar\sqsubseteq\katcontext$ of "variable declarations" and any "footprint" \metasig\ over \katvar\ (compare \cite{DW08,Wolter2021logics}). Up to now, we use, however, in our applications \cite{RLGRW14_JFAC,RRLW09_JLAP,Rutle10,RRLW10_JLAP_NWPT2009} only the category \katgraph\ of small (directed multi) graphs or categories \(\katgraph_{\graphTG}\) of typed graphs as \katcontext\ and we work with $\katvar=\katcontext$. Moreover, we have been using the term  "diagrammatic signature" instead of "footprint".
As statements we can use in DPF so-called "atomic $\metasig$-constraints" or arbitrary "first-order $\metasig$-constraints" lately introduced in \cite{Wolter2021logics}.
 
A \textbf{footprint} \metasig\ is given by set \(\setpred\) of "predicate symbols" and a map \(\flar{\arity}{\setpred}{\objectstype}\) assigning to each predicate symbol \(\predP\in\setpred\) its "arity" \(\arity(\predP)\in\objectstype\). An \textbf{atomic \metasig-constraint} \((\predP,\bindinga)\) on a context \contextK\ is given by a predicate symbol \(\predP\) and a morphism \(\bindinga:\arity(\predP)\to\contextK\).  As set \(\functorStm(\contextK)\) of all \textbf{\metasig-statements}, we consider here the set of all atomic \metasig-constraints \((\predP,\bindinga)\) on \contextK\ and for each context morphism \flar{\contextmorphisma}{\contextK}{\contextG} the map \flar{\functorStm(\contextmorphisma)}{\functorStm(\contextK)}{\functorStm(\contextG)} is simply defined by post-composition with \contextmorphisma:\, \(\functorStm(\contextmorphisma)(\predP,\bindinga)\isdef (\predP,\bindinga;\contextmorphisma)\). In DPF we call a pair $\sketchK=(\contextK,\setofstatementsK)$ with \(\setofstatementsK\subseteq\functorStm(\contextK)\) a \metasig-specification or a \textbf{\metasig-sketch}, respectively. 

Specification formalisms (modeling techniques) can be characterized by a certain choice of \katcontext, $\katvar\sqsubseteq\katcontext$ and \metasig. Footprints for the modeling techniques ``class diagrams'' and  ``relational data models'', e.g., are presented in \cite{Rutle10,RRLW10_JLAP_NWPT2009}. 

As running example, we use in this paper a simple but paradigmatic specification formalism, namely \textbf{Category Theory}. Categories are graphs equipped with a composition operation and an identity operation. So, we have \(\katcontext=\katgraph\) and we choose \katvar\ to be the full subcategory of \katgraph\ given by all finite graphs $\graphG=\graphlongG$ with \nodesG\ a finite subset of the set $\{v,\vertexv_1,$ $\vertexv_2,\ldots\}$ and \edgesG\ a finite subset of $\{e,e_1,e_2,\ldots\}$.
At the present stage of expansion, we do not include graph operations \cite{WolterDK18} in our footprints thus we have to formalize composition and identity by means of predicates. Therefore, the footprint \(\metasig_{CT}\) for the formalism ``Category Theory'' should declare two predicate symbols \featurecomp\ and \featureid. Besides this, we can, for example, include in \(\metasig_{CT}\) also the predicate symbols \featuremonic\ and \featurefinal\ to indicate the properties monomorphism and final object, respectively, with arities described in the following table.\\
\begin{center}
\vspace*{-3ex}
{\rm
\begin{tabular}{|p{9mm}|p{40mm}|p{9mm}|p{25mm}|}
\hline
\hspace{2ex} \predP& \hspace*{5mm} \textbf{Arity} $\arity(\predP)$ & \hspace{1mm} \predP& \hspace*{5mm} \textbf{Arity} $\arity(\predP)$ \\
\hline
  \hspace{1ex}\featurecomp & $\xymatrix{\vertexv_1 \ar[r]^{\edgee_1} \ar@/_.7pc/[rr]_{\edgee_3} & \vertexv_2 \ar[r]^{\edgee_2}  & \vertexv_3}$ 
& \hspace{2ex}\featureid & \hspace{5ex}$\xymatrix{v\ar@(r,d)[]^(.45){\edgee} }$  \\
\hline
  \hspace{1ex}\featuremonic & \hspace{5ex} $\xymatrix{\vertexv_1 \ar[r]^{\edgee}& \vertexv_2}$
& \hspace{1.5ex}\featurefinal & \hspace{5ex}$\xymatrix{\vertexv}$ \\
\hline
\end{tabular}}
\end{center}
As usual, we can visualize a graph morphism \flar{\varphi}{\graphA}{\graphG} for a finite graph \graphA\ by means of a visualization of the corresponding "\textbf{graph of assignments}" \(\graphA\varphi=(\nodesgen{\graphA\varphi},\edgesgen{\graphA\varphi},\sourcegen{\graphA\varphi},\targetgen{\graphA\varphi})\) with \(\nodesgen{\graphA\varphi}\isdef\{(\vertexv,\varphi_V(\vertexv))\mid\vertexv\in\nodesA\}\),  \(\edgesgen{\graphA\varphi}\isdef\{(\edgee,\varphi_E(\edgee))\mid\edgee\in\edgesA\}\) where  \(\sourcegen{\graphA\varphi}\) and \targetgen{\graphA\varphi} are defined for all \(\edgee\in\edgesA\) by \(\sourcegen{\graphA\varphi}(\edgee,\varphi_E(\edgee))=(\sourceA(\edgee),\varphi_V(\sourceA(\edgee)))\) and \(\targetgen{\graphA\varphi}(\edgee,\varphi_E(\edgee))=(\targetA(\edgee),\varphi_E(\targetA(\edgee)))\), respectively. 
For the graph \graphG\ below the graph morphism \(\gamma_1:\arity(\featurecomp)\to\graphG\) \\
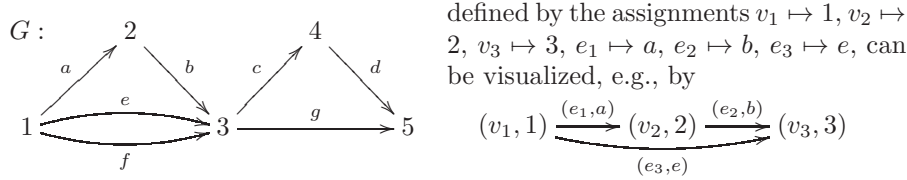
\begin{minipage}[c]{.5\linewidth}
\vspace*{-2ex}
$$\xymatrix{
\graphG:
& 2\ar[dr]^b && 4\ar[dr]^d \\
  1\ar[ur]^a\ar@/^/[rr]^e\ar@/_/[rr]_f
&& 3\ar[ur]^c\ar[rr]^g && 5
}$$
\end{minipage}
\begin{minipage}[c]{.5\linewidth}
\vspace*{.5ex}
defined by the assignments \(\vertexv_1\mapsto 1\), $\vertexv_2\mapsto 2$, $\vertexv_3\mapsto 3$, $e_1\mapsto a$, $e_2\mapsto b$, $e_3\mapsto e$, can be  visualized, e.g., by\\
\hspace*{2ex}$\xymatrix{(\vertexv_1,1) \ar[r]^{(e_1,a)} \ar@/_.7pc/[rr]_{(e_3,e)} & (\vertexv_2,2) \ar[r]^{(e_2,b)}  & (\vertexv_3,3)}$
\end{minipage}

If it is unambiguous,  we will use the shorthand notation \hspace*{1ex}$\xymatrix{1 \ar[r]^{a} \ar@/_.5pc/[rr]_{e} & 2 \ar[r]^{b}  & 3}$.

We extend the graph \graphG\ to a sample \(\metasig_{CT}\)-sketch \(\sketchG=(\graphG,\setofstatementsG)\) with \setofstatementsG\
a set of five statements in \contextG, i.e., five atomic \(\metasig_{CT}\)-con\-straints  \(\conditiong_1=(\featurecomp,\gamma_1)\), 
\(\conditiong_2=(\featurecomp,\xymatrix{1 \ar[r]^{a} \ar@/_.5pc/[rr]_{f} & 2 \ar[r]^{b} & 3})\),
\(\conditiong_3=(\featurecomp,\xymatrix{3 \ar[r]^{c} \ar@/_.5pc/[rr]_{g} & 4 \ar[r]^{d} & 5})\), 
\(\conditiong_4=(\featuremonic,\xymatrix{2 \ar[r]^{b}& 3})\),
\(\conditiong_5=(\featuremonic,\xymatrix{3 \ar[r]^{g}& 5})\).
If it is convenient and unambiguous, we will work with a visualization integrating graphs and atomic \metasig-constraints as in \cite{RLGRW14_JFAC,RRLW09_JLAP,Rutle10,RRLW10_JLAP_NWPT2009}. We can, for example, visualize graph \graphG\ with the four atomic \(\metasig_{CT}\)-constraints \\
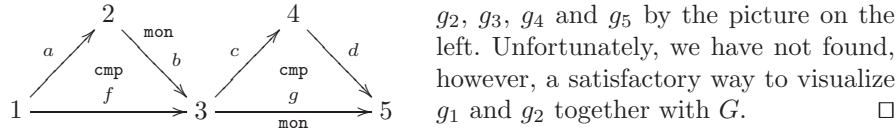
\begin{minipage}[c]{.5\linewidth}
\vspace*{-2ex}
$$\xymatrix{
&  2\ar[dr]^(.6)b\ar@{}[d]|(.6){\featurecomp} \ar@{}[dr]^(.3){\featuremonic} 
&& 4\ar[dr]^d \ar@{}[d]|(.6){\featurecomp}
\\
     1\ar[ur]^{a}\ar[rr]^f
&{}& 3\ar[ur]^c\ar[rr]^g_{\featuremonic}
&{}& 5
}$$
\end{minipage}
\begin{minipage}[c]{.5\linewidth}
\vspace*{.5ex}
$\conditiong_2$, $\conditiong_3$, $\conditiong_4$ and  $\conditiong_5$ by the picture on the left. Unfortunately, we have not found, however, a satisfactory way to visualize  \(\conditiong_1\) and  \(\conditiong_2\) together with \graphG.
\qed
\end{minipage}
\end{example}

\begin{example}[GraTra: Sketches]
\label{ex:gratra-sketches}
Traditionally, there is no explicit use of "statements" in the area of \textbf{graph transformations} thus sketches, in our sense, are just plain contexts where different kinds of graphs are chosen as contexts in the different approaches.
In \cite{HabelP09} \katcontext\ is a  category of directed, labeled multi graphs and \cite{BrugginkCHK11} restricts \katcontext\ to a category of finite directed, labeled multi graphs. In contrast, \cite{Rensink04} works with directed, labeled simple graphs in the sense, that parallel edges with the same label are not allowed. \cite{Kosiol0TZ20} uses as \katcontext\  a  category \(\katgraph_{\graphTG}\) of directed, labeled multi graphs typed over a graph \graphTG.

To a certain extend we can, however, interpret the transition from graphs to labeled/typed graphs as the utilization of rudimentary forms of "statements"
where the choice of label alphabets or type graphs \graphTG, respectively, corresponds to the choice of footprints in DPF. The encoding of binary relations by means of labeled edges in \cite{Rensink04} makes this analogy apparent. Within DPF we can reconstruct the concept of graph in \cite{Rensink04} in the following way: \katcontext\ is the subcategory of \katset\ given by all subsets of a "countable universe of nodes \textsf{Node}" and $\katvar\sqsubset\katcontext$ has a two-element set \(\{x_1,x_2\}\subset\mathsf{Node}\) as its only object. The footprint $\metasig_R$ is given by a "countable universe \textsf{Rel}" of predicate symbols with \(\arity(\predP)=\{x_1,x_2\}\) for all \(\predP\in\mathsf{Rel}\). An atomic $\metasig_R$-constraints on a context \(\contextK\subseteq\mathsf{Node}\) becomes, in such a way, a pair \((\predP,\bindinga)\) with \(\predP\in\mathsf{Rel}\) and  \(\bindinga:\{x_1,x_2\}\to\contextK\) a map. Relying on the isomorphism between the Cartesian product  \(\contextK\times\contextK\) and the set \(\contextK^{\{x_1,x_2\}}\) of maps, it is easy to check that the category \textsf{Graph} in \cite{Rensink04} is isomorphic to the non-adhesive category of all $\metasig_R$-sketches.
\qed
\end{example}
\section{First-order sketch conditions and constraints}
\label{sec:first-order-sketch-conditions}
Generalizing different variants of "graph conditions" \cite{BrugginkCHK11,EEPT06,HabelP09,Kosiol0TZ20,Rensink04} as well as "universal conditions" and "negative universal conditions" in DPF \cite{Rutle10,RRLW10_JLAP_NWPT2009}, we introduce general first-order sketch conditions in a redundant manner,
in the sense, that we introduce, for example, as well existential as universal quantification.  We define first-order sketch conditions analogously to "first-order expressions" in \cite{Wolter2021logics}. 

%

\begin{definition}[Sketch conditions: Syntax]
\label{def:sketch-conditions:syntax}
For any category \katcontext\ of contexts and any statement functor \flar{\functorStm}{\katcontext}{\katset} we define inductively the set \setofsketchcond{\contextK} of all \textbf{first-order sketch conditions over context \contextK} 
by the following rules: 
\begin{enumerate}
\item Statements: \;$\functorStm(\contextK)\subset\setofsketchcond{\contextK}$ \; for any context \contextK.
\item True: \;$\true\in\setofsketchcond{\contextK}$ \; for any context \contextK.
\item False: \;$\false\in\setofsketchcond{\contextK}$ \; for any context \contextK.
\item Conjunction: $\bigwedge\setC\in\setofsketchcond{\contextK}$ \; for any set \(\setC\subset\setofsketchcond{\contextK}\) of conditions over $\contextK$.
\item Disjunction: $\bigvee\setC\in\setofsketchcond{\contextK}$ \; for any set \(\setC\subset\setofsketchcond{\contextK}\) of conditions over $\contextK$.
\item Negation: $\neg\condition\in\setofsketchcond{\contextK}$ for any condition $\condition\in\setofsketchcond{\contextK}$.
\item Guarded existential quantification:
$\guardedexistential{\conditiona}{\contextmorphisma}{\contextM}{\conditionb} \in\setofsketchcond{\contextK}$ for any $\conditiona\in\setofsketchcond{\contextK}$, $\conditionb\in\setofsketchcond{\contextM}$ and any morphism $\contextmorphisma:\contextK\to\contextM$ in \katcontext.
\item Guarded universal quantification:   $\guardeduniversal{\conditiona}{\contextmorphisma}{\contextM}{\conditionb}\in\setofsketchcond{\contextK}$ for any $\conditiona\in\setofsketchcond{\contextK}$, $\conditionb\in\setofsketchcond{\contextM}$ and any morphism $\contextmorphisma:\contextK\to\contextM$ in \katcontext. 
\end{enumerate}
We write also \(\contextK\contextof\condition\), instead of \(\condition\in\setofsketchcond{\contextK}\), and call \contextK\ the \textbf{context of \condition}.
\end{definition}
\begin{remark}[Notations]
\label{rem:notations}
Analogously to \cite{BrugginkCHK11,HabelP09,Rensink04}, we allow non-monic morphisms $\contextmorphisma:\contextK\to\contextM$ in order to express identifications (compare Example \ref{ex:sketch-constraints-category-theory}). 

In case \(\contextK=\contextM\) and \(\contextmorphisma=\identityof{\contextK}\), both quantifications become obsolet and we get a kind of "propositional implication" just written as \propimplication{\conditiona}{\conditionb}. 

In case \(\conditiona=\true\), quantification is unguarded and we just write \uncondexists{\contextmorphisma}{\contextM}{\conditionb} and \uncondforall{\contextmorphisma}{\contextM}{\conditionb} for the existential quantification \guardedexistential{\true}{\contextmorphisma}{\contextM}{\conditionb} and universal quantification \guardeduniversal{\true}{\contextmorphisma}{\contextM}{\conditionb}, respectively.

In  cases, where \katcontext\ is set-based, as \katset, \katgraph\ or \(\katgraph_{\graphTG}\), for example, contexts \contextK\ are constituted by single entities thus we can talk about individual "variables". \(\contextK\contextof\condition\) tells us then, especially, that \contextK\ comprises all "free variables" in  \condition. If \initialobject\ is  an initial object in \katcontext, \(\initialobject\contextof\condition\) means that all "variables" in \condition\ are bounded and we call \condition\ a \textbf{closed sketch condition}. For set-based categories \katcontext\ we may also just drop \contextmorphisma\ if $\contextmorphisma=in_{\contextK,\contextM}:\contextK\hookrightarrow\contextM$ is an "inclusion morphism".
%
\qed
\end{remark}
\begin{remark}[GraTra: Conditions]
\label{rem:gratra-conditions}
Sketch conditions with only "quantification free" guards   \conditiona\ in all guarded quantifications are   "tree-like" conditions analogously to the conditions in  \cite{BrugginkCHK11,HabelP09,Kosiol0TZ20,Rensink04}. They can be seen as a generalizing modification of the "Q(uantifier)-trees" of the "language of diagrams" in \cite{FreydS1990}.


All approaches \cite{BrugginkCHK11,HabelP09,Kosiol0TZ20} rely on unguarded quantifications, i.e., on guards $\conditiona=\true$. In \cite{HabelP09} and \cite{Kosiol0TZ20} the unguarded variant \uncondexists{\contextmorphisma}{\contextM}{\conditionb} of existential quantification is used and  \uncondforall{\contextmorphisma}{\contextM}{\conditionb} is encoded by $\neg\uncondexists{\contextmorphisma}{\contextM}{\neg\conditionb}$.
In \cite{BrugginkCHK11} the symbols "$\exists$" and "$\forall$" are used in a bit unconventional, but consistent, way: In view of Definition \ref{def:sketch-conditions:syntax}, "$\exists$" combines "disjunction and existential quantification" while  "$\forall$" combines conjunction and universal quantification. The conditions in \cite{BrugginkCHK11} correspond to sketch conditions that can be generated by a rule like: $\bigvee\{\uncondexists{\contextmorphisma_i}{\contextM_i}{\condition_i}\mid i\in\setI\},\bigwedge\{\uncondforall{\contextmorphisma_i}{\contextM_i}{\condition_i}\mid i\in\setI\}\in\setofsketchcond{\contextK}$ for any family \(\{\contextmorphisma_i:\contextK\to\contextM_i\mid i\in\setI\}\) of context morphisms and any conditions \(\condition_i\in\setofsketchcond{\contextM_i}\), \(i\in\setI\). \true\ is encoded by the empty conjunction \(\bigwedge\emptyset\) and \false\ by the empty disjunction \(\bigvee\emptyset\), respectively. 
\qed
\end{remark}

Generalizing the traditional ways \cite{BrugginkCHK11,EEPT06,HabelP09,Kosiol0TZ20,Rensink04} to define a satisfaction relation between graph morphisms and graph conditions, we can define a satisfaction relation   between context morphisms and sketch conditions.

\begin{definition}[Sketch conditions: Satisfaction]
\label{def:sketch-conditions:satisfaction}
We define a satisfaction relation \morphismsatisfies{\contextmorphismt}{\sketchG}{\condition} between context morphisms \(\contextmorphismt:\contextK\to\contextG\) and sketch conditions \(\condition\in\setofsketchcond{\contextK}\) relative to a sketch \(\sketchG=(\contextG,\setofstatementsG)\) as follows:
\begin{enumerate}
\item Statement:  For all $\statement\in\functorStm(\contextK)\subset\setofsketchcond{\contextK}$: \, \morphismsatisfies{\contextmorphismt}{\sketchG}{\statement} \, iff \, \(\contextmorphismt(\statement)\in\setofstatementsG\).
\item True: \, \morphismsatisfies{\contextmorphismt}{\sketchG}{\true}
\item False: \, \morphismsatisfiesnot{\contextmorphismt}{\sketchG}{\false}
\item Conjunction: \, \morphismsatisfies{\contextmorphismt}{\sketchG}{\bigwedge\setC} \, iff \, \morphismsatisfies{\contextmorphismt}{\sketchG}{\condition}\; for every \(\condition\in\setC\).
\item Disjunction: \, \morphismsatisfies{\contextmorphismt}{\sketchG}{\bigvee\setC} \, iff \, \morphismsatisfies{\contextmorphismt}{\sketchG}{\condition}\; for some \(\condition\in\setC\).
\item Negation: \, \morphismsatisfies{\contextmorphismt}{\sketchG}{\neg\condition} \, iff \, \morphismsatisfiesnot{\contextmorphismt}{\sketchG}{\condition}.
\item Guarded existential quantification:
\, \morphismsatisfies{\contextmorphismt}{\sketchG}{\guardedexistential{\conditiona}{\contextmorphisma}{\contextM}{\conditionb}}  iff  
\morphismsatisfies{\contextmorphismt}{\sketchG}{\conditiona} implies that there is a morphism \(\contextmorphismr:\contextM\to\contextG\) such that \(\contextmorphisma;\contextmorphismr=\contextmorphismt\) and \morphismsatisfies{\contextmorphismr}{\sketchG}{\conditionb}.
\item Guarded universal quantification:  
\, \morphismsatisfies{\contextmorphismt}{\sketchG}{\guardeduniversal{\conditiona}{\contextmorphisma}{\contextM}{\conditionb}} \, iff \, 
\morphismsatisfies{\contextmorphismt}{\sketchG}{\conditiona} implies that for all morphisms \(\contextmorphismr:\contextM\to\contextG\) with \(\contextmorphisma;\contextmorphismr=\contextmorphismt\) it holds \morphismsatisfies{\contextmorphismr}{\sketchG}{\conditionb}.
\end{enumerate}
\end{definition}
Of course we can restrict, if necessary, the morphism  \contextmorphismt\ and/or the morphisms \contextmorphismr\ to a certain classes of morphisms like monomorphisms, for example.

The satisfaction of graph/sketch conditions by a morphism is a powerful and practical useful tool to control the application of transformation rules. This is extensively demonstrated and validated in the Graph Transformation literature as in \cite{BrugginkCHK11,EEPT06,HabelP09,Kosiol0TZ20,Rensink04}, for example. In DPF we used until now only non-nested negative application conditions to control the application of non-deleting model transformation rules \cite{Rutle10,RRLW10_JLAP_NWPT2009}. The paper paves the way for utilizing arbitrary first-order conditions to control "model transformations" in DPF. In this paper we will, however, not explore this promising direction of applying first-order sketch conditions. We rather concentrate on two other aspects of diagrammatic modeling techniques 
- namely "syntactic structure" of models and
"deducing information from and  reason about models" in a diagrammatic manner. 

Developing and applying DPF, we realized that "typing mechanisms" are not powerful enough to formalize all relevant restrictions concerning the syntactic structure of models. To overcome this deficiency we introduced "universal constraints" and "negative universal constraints" \cite{Rutle10,RRLW10_JLAP_NWPT2009} in analogy to the non-nested graph constraints in \cite{EEPT06}. Fortunately, sketch conditions and their satisfaction, as defined in Definition \ref{def:sketch-conditions:satisfaction}, give us now also more powerful general first-order sketch constraints at hand to describe the syntactic structure of models. 
The simple, but crucial, observation is that the assertion \morphismsatisfies{\contextmorphismt}{\sketchG}{\condition} can be interpreted as well as an assertion concerning the structure of \sketchG.
\begin{definition}[Sketch constraints]
\label{def:sketch-contraints}
A \textbf{sketch constraint} \((\condition,\contextmorphismt)\) on context \contextG\ is given by a sketch condition \(\contextK\contextof\condition\)
and a context morphism \(\contextmorphismt:\contextK\to\contextG\).

A sketch \sketchG\ with underlying context \contextG, i.e., \(\sketchG=(\contextG,\setofstatementsG)\), \textbf{satisfies} the constraint \((\condition,\contextmorphismt)\), \(\sketchsatisfies{\sketchG}{\condition}{\contextmorphismt}\) in symbols, if, and only if, \morphismsatisfies{\contextmorphismt}{\sketchG}{\condition}.
\end{definition}
If the sketch condition \condition\ does not contain any statements, as it usually the case in the area of graph transformations (compare Example \ref{ex:gratra-sketches}), \(\sketchsatisfies{\sketchG}{\condition}{\contextmorphismt}\) is just an assertion about the structure of the context \contextG. In all other cases, \(\sketchsatisfies{\sketchG}{\condition}{\contextmorphismt}\) tells us also something about the presence or non-presence of statements as well as the relations between the statements in \sketchG.  

Due to rule "\textit{Statement}" all statements reappear as conditions. The following simple Corollary illustrates that the  requirement for sketch morphisms to preserves statements "on the nose" encodes a structural constraint on the target. 
\begin{corollary}[Sketch morphism vs. sketch constraint]
\label{coro:sketch-morphism-vs-sketch-constraint}
A context morphism $\contextmorphisma:\contextK\to\contextG$ constitutes a morphism $\contextmorphisma:\sketchK\to\sketchG$ between two sketches $\sketchK=(\contextK,\setofstatementsK)$ and $\sketchG=(\contextG,\setofstatementsG)$ if, and only if, \sketchsatisfies{\sketchG}{\bigwedge\setofstatementsK}{\contextmorphisma}.
\end{corollary}
\begin{remark}[Global constraints]
\label{rem:global-constraints}
A constraint \((\condition,\contextmorphismt)\) is, in general, only  a \textbf{"local constraint"}, in the sense, that it constraints the structure of \sketchG\ "around the image" of \contextK\ w.r.t.\  \contextmorphismt.
If \katcontext\ has an initial object \initialobject, any closed condition \(\initialobject\contextof\condition\) 
gives rise to a unique sketch constraint \((\condition,\initialmorphism{\contextG})\) with \(\initialmorphism{\contextG}:\initialobject\to\contextG\) the initial morphism into \contextG. \((\condition,\initialmorphism{\contextG})\) is a \textbf{"global constraint"}, in the sense that the statement  \(\sketchsatisfies{\sketchG}{\condition}{\initialmorphism{\contextG}}\) can be seen as a statement concerning the overall structure of \sketchG. "Local constraints" are usually not considered in the literature \cite{HabelP09,Kosiol0TZ20}.
\qed
\end{remark}

\begin{example}[Sketch constraints: Category theory]
\label{ex:sketch-constraints-category-theory}
All requirements that turn a  \(\metasig_{CT}\)-sketch (see Example \ref{ex:dpf-sketches}), into (a presentation of) a category give rise to local constraints and corresponding global constraints. 
%
Local constraints that composition is defined for a certain pair of  edges, for example, can be formalized by condition \(\conditionct_1\) where we rely here on the notational conventions in Remark \ref{rem:notations}:\\
\hspace*{7ex}\(\conditionct_1\isdef\;
\xymatrix{\vertexv_2\ar[dr]^{\edgee_2} \\ \vertexv_1 \ar[u]^{\edgee_1} & \vertexv_3}\contextof 
\exists(\xymatrix{\vertexv_2\ar[dr]^{\edgee_2} \\ \vertexv_1 \ar[r]^(.4){\edgee_3}\ar[u]^{\edgee_1} & \vertexv_3}:
\xymatrix{\vertexv_2\ar[dr]^{\edgee_2}\ar@{}[dr]_(.5){\featurecomp} \\ \vertexv_1 \ar[r]^(.5){\edgee_3}\ar[u]^{\edgee_1} & \vertexv_3}  )
\)\\
Universal quantification transforms this condition into a closed condition \(\conditionct_2\) that composition is always defined where \initialobject\ denotes here the empty graph:\\
\hspace*{7ex}
\(\conditionct_2\isdef\;
\initialobject\contextof 
\forall(
\xymatrix{\vertexv_2\ar[dr]^{\edgee_2} \\ \vertexv_1 \ar[u]^{\edgee_1} & \vertexv_3}:
\exists(\xymatrix{\vertexv_2\ar[dr]^{\edgee_2} \\ \vertexv_1 \ar[r]^(.4){\edgee_3}\ar[u]^{\edgee_1} & \vertexv_3}:
\xymatrix{\vertexv_2\ar[dr]^{\edgee_2}\ar@{}[dr]_(.5){\featurecomp} \\ \vertexv_1 \ar[r]^(.5){\edgee_3}\ar[u]^{\edgee_1} & \vertexv_3}  
) 
)
\)
\\
For the sample sketch \(\sketchG=(\contextG,\setofstatementsG)\) in Example \ref{ex:dpf-sketches}, we do have \(\sketchsatisfies{\sketchG}{\conditionct_1}{\contextmorphismt_1}\), with \(\contextmorphismt_1\) given by the assignments \(\edgee_1\mapsto a,\edgee_2\mapsto b\), but \(\sketchsatisfiesnot{\sketchG}{\conditionct_1}{\contextmorphismt_2}\),  with \(\contextmorphismt_2\) given by \(\edgee_1\mapsto b,\edgee_2\mapsto c\), thus \(\sketchsatisfiesnot{\sketchG}{\conditionct_2}{\initialmorphism{\contextG}}\).

Global constraints imposing uniqueness of composition, independent of the existence of composition, can be formulated by the  closed condition \(\conditionct_3\):\\
\hspace*{2ex}
\(
\initialobject\contextof 
\forall(\hspace*{-1ex}
\xymatrix{\vertexv_2\ar[dr]^{\edgee_2} \\ \vertexv_1\ar@/_.2pc/[r]_(.4){\edgee_4}\ar@/^.2pc/[r]^(.4){\edgee_3} \ar[u]^{\edgee_1} & \vertexv_3}\hspace*{-2ex}\hspace*{-1ex}:
(\bigwedge
\{\hspace*{-1ex}\xymatrix{\vertexv_2\ar[dr]^{\edgee_2}\ar@{}[dr]_(.5){\featurecomp} \\ \vertexv_1 \ar[r]^(.5){\edgee_3}\ar[u]^{\edgee_1} & \vertexv_3}\hspace*{-2ex} ,
\hspace*{-.5ex}\xymatrix{\vertexv_2\ar[dr]^{\edgee_2}\ar@{}[dr]_(.5){\featurecomp} \\ \vertexv_1 \ar[r]^(.5){\edgee_4}\ar[u]^{\edgee_1} & \vertexv_3} 
\hspace*{-1ex}\}\to
\exists(\contextmorphisma,\hspace*{-1ex}\xymatrix{\vertexv_2\ar[dr]^{\edgee_2}&:\true \\ \vertexv_1 \ar[r]^(.4){e}\ar[u]^{\edgee_1} & \vertexv_3}
) ) )
\)
\\
Note, that existential quantification is guarded this time. 
\contextmorphisma\ simply maps $\edgee_3$ and $\edgee_4$ to \(\edgee\).
\sketchG\ doesn't satisfies the constraint \((\conditionct_3,\initialmorphism{\contextG})\) but would satisfy it if we delete edge "$f$", for example.
The remaining requirements -- existence and uniqueness of identities, identity laws and associativity law -- can be expressed analogously. 

Besides formalizing the "laws of a category", we would, however, also like to take advantage of our knowledge about the properties of the predicates in \(\metasig_{CT}\). Or to put it the other way around: We would like to formulate requirements that any intended semantics of the  predicates in \(\metasig_{CT}\) has to comply with. For example, we can require that for a final object all outgoing morphisms are monic:\\
\hspace*{7ex}\(\conditionct_4\isdef\;\initialobject\contextof \forall(\vertexv:(\vertexv^\featurefinal\longrightarrow
\forall(
\vertexv\stackrel{\edgee}{\longrightarrow}\vertexv_1: 
\xymatrix{\vertexv \ar[r]^{e}_{\featuremonic} & \vertexv_1} )))
\)
\\
We can require that monomorphisms are closed under composition:
\\
\(\hspace*{7ex}\conditionct_5\isdef\;
\initialobject\contextof 
\forall(\hspace*{-1ex}
\xymatrix{\vertexv_2\ar[dr]^{\edgee_2} \\ \vertexv_1\ar[r]^(.4){\edgee_3} \ar[u]^{\edgee_1} & \vertexv_3}\hspace*{-1ex}:
(\bigwedge
\{
\xymatrix{\vertexv_2\ar[dr]^[.4]{\edgee_2}\ar@{}[dr]_(.5){\featurecomp}^(.7){\featuremonic} \\ \vertexv_1 \ar[r]^(.5){\edgee_3}\ar[u]^(.6){\edgee_1}\ar@{}[u]^(.3){\featuremonic} & \vertexv_3}
\hspace*{-1ex}\}\longrightarrow
\xymatrix{\vertexv_2\ar[dr]^{\edgee_2} \\ \vertexv_1 \ar[r]^(.4){\edgee_3}\ar@{}[r]_(.4){\featuremonic}\ar[u]^{\edgee_1} & \vertexv_3}
) ) 
\)
\\
Note, that we use \(\bigwedge\{\cdots\}\) because the single triangle between the curly brackets visualizes, actually, three \(\metasig_{CT}\)-statements (atomic \(\metasig_{CT}\)-constraints). We can also express our knowledge concerning the decomposition of monomorphisms:
\\
\(\hspace*{7ex}\conditionct_6\isdef\;
\initialobject\contextof 
\forall(\hspace*{-1ex}
\xymatrix{\vertexv_2\ar[dr]^{\edgee_2} \\ \vertexv_1\ar[r]^(.4){\edgee_3} \ar[u]^{\edgee_1} & \vertexv_3}\hspace*{-1ex}:
(\bigwedge
\{
\xymatrix{\vertexv_2\ar[dr]^[.4]{\edgee_2}\ar@{}[dr]_(.5){\featurecomp} \\ \vertexv_1 \ar[r]^(.4){\edgee_3}\ar@{}[r]_(.4){\featuremonic}\ar[u]^(.4){\edgee_1} & \vertexv_3}
\hspace*{-1ex}\}\longrightarrow
\xymatrix{\vertexv_2\ar[dr]^(.5){\edgee_2} \\ \vertexv_1 \ar[r]^(.4){\edgee_3}\ar[u]^(.53){\edgee_1}\ar@{}[u]^(.3){\featuremonic} & \vertexv_3}
) ) 
\)
\\
\(\sketchsatisfies{\sketchG}{\conditionct_5}{\initialmorphism{\contextG}}\) simply because there is no match in \contextG\ of the triangular context  in \(\conditionct_5\) satisfying the premise of the implication in \(\conditionct_5\). In contrast, \(\sketchsatisfiesnot{\sketchG}{\conditionct_6}{\initialmorphism{\contextG}}\) with the only counterexample given by the assignments \(e_1\mapsto c,e_2\mapsto d,e_3\mapsto g \).

That concepts and constructions are defined by \textbf{universal properties} is the crucial characteristic of Category Theory as a specification formalism. The concept "monomorphism", for example, is defined by the universal property:
\\
\(\conditionct_7\isdef
\xymatrix{\vertexv_1\ar[d]^{e}  \\ \vertexv_2}
\hspace*{-1.5ex}
\contextof 
\forall(\hspace*{-2ex}
\xymatrix{&\vertexv_1\ar[d]^{e} \\ \vertexv_3\ar[r]^(.5){\edgee_3}\ar@/_.2pc/[ur]_(.56){\edgee_2}\ar@/^.2pc/[ur]^(.4){\edgee_1} & \vertexv_2}\hspace*{-1ex}:\hspace*{-.5ex}
(\bigwedge
\{\hspace*{-3.5ex}
\xymatrix{&\vertexv_1\ar[d]^{\edgee}\ar@{}[dl]^(.45){\featurecomp} \\ \vertexv_3\ar[r]^(.5){\edgee_3}\ar[ur]^(.4){\edgee_1} & \vertexv_2},
\hspace*{-1.5ex} 
\xymatrix{&\vertexv_1\ar[d]^{\edgee}\ar@{}[dl]^(.45){\featurecomp} \\ \vertexv_3\ar[r]^(.5){\edgee_3}\ar[ur]^(.4){\edgee_2} & \vertexv_2}
\hspace*{-1ex}\}
\longrightarrow
\exists(\contextmorphisma,\hspace*{-2ex}
\xymatrix{&\vertexv_1\ar[d]^{\edgee} \\ \vertexv_3\ar[r]^(.5){\edgee_3}\ar[ur]^(.4){\edgee_4} & \vertexv_2}:\true
)) )
\)
\\
where \contextmorphisma\ maps $\edgee_1$ and $\edgee_2$ to \(\edgee_4\).
In most cases, however, a universal property is the conjunction of a universally quantified existence assertion and a universally quantified uniqueness assertion. The concept "final object", for example, is defined by the universal property \(\conditionct_8\):\\
\(\hspace*{4ex}\conditionct_8\isdef
\vertexv\contextof 
\bigwedge\{
\forall(\vertexv_1\;\; \vertexv:\exists(\vertexv_1\stackrel{\edgee}{\longrightarrow}\vertexv
: \true )),
\forall(\xymatrix{\vertexv_1\ar@/_.2pc/[r]_(.5){\edgee_2}\ar@/^.2pc/[r]^(.5){\edgee_1}& \vertexv}:
\exists(\contextmorphisma,
\vertexv_1\stackrel{\edgee}{\longrightarrow}\vertexv: \true ))
\}
\)
\\
where \contextmorphisma\ maps $\edgee_1$ and $\edgee_2$ to $\edgee$. A procedure, useful and needed to reason about and to work with \(\metasig_{CT}\)-sketches, is the replacement of \featurefinal-statements and \featuremonic-statements, for example, by the corresponding universal properties.
Replacing the \featurefinal-statement \(\vertexv^\featurefinal\) in condition \(\conditionct_4\) by \(\conditionct_8\), for example, is unproblematic since \(\conditionct_8\) is a condition over $\arity(\featurefinal)=\vertexv$. We get the condition \(\conditionct_4'\):
\\
\(
\initialobject\contextof 
\hspace*{-.5ex}
\forall(\vertexv:
(\bigwedge\{
\forall(\vertexv_1\;\; \vertexv:\exists(\vertexv_1\stackrel{\edgee}{\longrightarrow}\vertexv
: \true )),
\forall(\xymatrix{\vertexv_1\ar@/_.2pc/[r]_(.5){\edgee_2}\ar@/^.2pc/[r]^(.5){\edgee_1}& \vertexv}:
\exists(\contextmorphisma,
\vertexv_1\stackrel{\edgee}{\longrightarrow}\vertexv: \true ))
\}
\\
{}\hfill\longrightarrow
\forall(
\vertexv\stackrel{\edgee}{\longrightarrow}\vertexv_1: 
\xymatrix{\vertexv \ar[r]^{e}_{\featuremonic} & \vertexv_1} ))
\)
\\
Note, that we get here a guard with nested quantifiers!

Unfolding then the \featuremonic-statement in \(\conditionct_4'\) by the definition \(\conditionct_7\) of the property "monic" is also unproblematic since the context of \(\conditionct_7\) is isomorphic to the context $\vertexv\stackrel{\edgee}{\longrightarrow}\vertexv_1$ of the statement $\xymatrix{\vertexv \ar[r]^{e}_{\featuremonic} & \vertexv_1}$ in \(\conditionct_4'\).
If we rename $\vertexv_1\stackrel{\edgee}{\longrightarrow}\vertexv_2$ in \(\conditionct_7\) by $\vertexv\stackrel{\edgee}{\longrightarrow}\vertexv_1$,  we can just replace the statement $\xymatrix{\vertexv \ar[r]^{e}_{\featuremonic} & \vertexv_1}$ in \(\conditionct_4'\) by the corresponding renamed variant of \(\conditionct_7\).
To unfold, however, the \featuremonic-statements in conditions \(\conditionct_5\) and \(\conditionct_6\), for example, we have to translate condition \(\conditionct_7\) into a condition over the corresponding bigger contexts of these \featuremonic-statements.
\qed
\end{example}
The translation of conditions is related to the shift operation in \cite{BrugginkCHK11}.
\begin{proposition}
\label{prop:translation-of-conditions}
We assume that \katcontext\ has pushouts. For an arbitrary but fixed choice of pushouts in \katcontext\ we can define for any context morphism $\contextmorphismc:\contextK\to\contextH$ a \textbf{translation map} $\overline{\contextmorphismc}:\setofsketchcond{\contextK}\to\setofsketchcond{\contextH}$ such that \(in_\contextK;\overline{\contextmorphismc}=\functorStm(\contextmorphismc);in_\contextH\) for the inclusion maps \(in_\contextK:\functorStm(\contextK)\to\setofsketchcond{\contextK}\) and \(in_\contextG:\functorStm(\contextH)\to\setofsketchcond{\contextH}\).
\begin{enumerate}
	\item Statements: \(\overline{\contextmorphismc}(\statement)\isdef\contextmorphismc(\statement)\) \; for all \(\statement\in\functorStm(\contextK)\).
	\item True: \;$\overline{\contextmorphismc}(\true)\isdef\true$.
	\item False: \;$\overline{\contextmorphismc}(\false)\isdef\false$.
	\item Conjunction: $\overline{\contextmorphismc}(\bigwedge\setC)\isdef\bigwedge\{\overline{\contextmorphismc}(\condition)\mid \condition\in\setC\}$.
	\item Disjunction: $\overline{\contextmorphismc}(\bigvee\setC)\isdef\bigvee\{\overline{\contextmorphismc}(\condition)\mid \condition\in\setC\}$.
	\item Negation: $\overline{\contextmorphismc}(\neg\condition)\isdef\neg\overline{\contextmorphismc}(\condition)$.
	\item Guarded existential quantification: Let \(\contextH\stackrel{\contextmorphisma^*}{\rightarrow}\contextM_\contextmorphismc\stackrel{\contextmorphismc^*}{\leftarrow}\contextM\) be the chosen pushout of \(\contextH\stackrel{\contextmorphismc}{\leftarrow}\contextK\stackrel{\contextmorphisma}{\rightarrow}\contextM\):\;
	$\overline{\contextmorphismc}\guardedexistential{\conditiona}{\contextmorphisma}{\contextM}{\conditionb}\isdef\guardedexistential{\overline{\contextmorphismc}(\conditiona)}{\contextmorphisma^*}{\contextM_\contextmorphismc}{\overline{\contextmorphismc^*}(\conditionb)}$.
	\item Guarded universal quantification:  Let \(\contextH\stackrel{\contextmorphisma^*}{\rightarrow}\contextM_\contextmorphismc\stackrel{\contextmorphismc^*}{\leftarrow}\contextM\) be the chosen pushout of \(\contextH\stackrel{\contextmorphismc}{\leftarrow}\contextK\stackrel{\contextmorphisma}{\rightarrow}\contextM\):\;
	$\overline{\contextmorphismc}\guardeduniversal{\conditiona}{\contextmorphisma}{\contextM}{\conditionb}\isdef\guardeduniversal{\overline{\contextmorphismc}(\conditiona)}{\contextmorphisma^*}{\contextM_\contextmorphismc}{\overline{\contextmorphismc^*}(\conditionb)}$.
\end{enumerate}
\end{proposition}
\begin{remark}[Chosen pushouts]
\label{rem:chosen-pushouts}
In case \contextmorphismc\ is an isomorphism, the best choice for a pushout is, of course, the cospan \(\contextH\stackrel{\contextmorphismc^{-1}:\contextmorphisma}{\longrightarrow}\contextM\stackrel{id_\contextM}{\longleftarrow}\contextM\).
\qed
\end{remark}
The translation \(\overline{\contextmorphismc}(\conditionct_7)\) of the universal property \(\conditionct_7\) of monomorphisms in Example \ref{ex:sketch-constraints-category-theory} along the unique graph morphism \(\contextmorphismc:(\vertexv_1\stackrel{\edgee}{\longrightarrow}\vertexv_2)\to\xymatrix{v\ar@(ur,dr)[]^(.45){\edgee} }\) gives us, for example, a definition of monic loops at hand. 

Note, that the assignments \(\contextmorphismc\mapsto\overline{\contextmorphismc}\) define only a pseudofunctor \(\functorSC:\katcontext\to\katset\) since, in general, the composition of chosen pushouts does not result in a chosen pushout. 
This may be a hint to develop a future deduction calculus for sketch constraints rather in a fibred setting (compare \cite{WolterMH2020})?
%
We close this section with a short analysis of the structure of universal properties defining (co)limits.
\begin{remark}[Sketch constraints: (Co)limits]
\label{rem:sketch-constraints-colimits} 
The universal property defining a (co)limit is the conjunction of two assertions -- existence of mediators and uniqueness of mediators. We can express those assertions by \(\metasig_{CT}\)-sketch conditions with the following structure (compare the definition of final objects in Example \ref{ex:sketch-constraints-category-theory}):\\
\(\hspace*{4ex}exist_\graphI\isdef
C_\graphI\contextof 
\forall(C_\graphI+C'_\graphI:(\conditionct_1\longrightarrow\exists(C_\graphI\stackrel{\to}{+}C'_\graphI:\conditionct_2 )))\\
\hspace*{4ex}unique_\graphI\isdef
C_\graphI\contextof \forall(C_\graphI\stackrel{\Rightarrow}{+}C'_\graphI:(\conditionct_3\longrightarrow\exists(\contextmorphisma,C_\graphI\stackrel{\to}{+}C'_\graphI:\true )))
\)\\
\graphI\ is the shape graph of the (co)limit. \(C_\graphI\) adds to \graphI\ the shape of a (co)cone with base \graphI\ while \(C_\graphI+C'_\graphI\) extends \(C_\graphI\) with the shape of a second (co)cone with base \graphI. \(\conditionct_1\) is the conjunction of all \featurecomp-statements in  \(C_\graphI+C'_\graphI\) turning both (co)cones into commutative ones. \(C_\graphI\stackrel{\to}{+}C'_\graphI\) extends \(C_\graphI+C'_\graphI\) by a single mediator  while \(\conditionct_2\) is the conjunction of \featurecomp-statements in \(C_\graphI\stackrel{\to}{+}C'_\graphI\) expressing the commutativity requirements for the mediator. \(C_\graphI\stackrel{\Rightarrow}{+}C'_\graphI\) extends \(C_\graphI+C'_\graphI\) by two parallel mediators and \(\conditionct_3\) is the conjunction of \featurecomp-statements in \(C_\graphI\stackrel{\Rightarrow}{+}C'_\graphI\) expressing the commutativity requirements for both mediators. \(\contextmorphisma:C_\graphI\stackrel{\Rightarrow}{+}C'_\graphI\longrightarrow C_\graphI\stackrel{\to}{+}C'_\graphI\) simply identifies the two mediators in \(C_\graphI\stackrel{\Rightarrow}{+}C'_\graphI\).
\qed
\end{remark}

\section{Sketch morphisms, constraints and deduction}
\label{sec:sketch-rules}

In this section we present vital observations, insights, concepts and ideas to establish a basis for the future further development of the "logic dimension" of DPF based on the new concepts and results presented in this paper.

\noindent
\textbf{Constraints in DPF at present:}
Following \cite{Mak97} and in analogy to \cite{EEPT06}, we use in DPF until now, instead of sketch constraints in the sense of Definition \ref{def:sketch-contraints}, only plain sketch morphisms $\contextmorphisma:\sketchL\to\sketchR$ and call them "(positive) universal constraints" or "negative universal constraints", respectively \cite{Rutle10,RRLW10_JLAP_NWPT2009}. We define that a sketch \sketchG\ satisfies the "universal constraint" $\contextmorphisma:\sketchL\to\sketchR$ if, and only if, for any sketch morphism \(\contextmorphismt:\sketchL\to\sketchG\) there is a sketch morphism \(\contextmorphismr:\sketchR\to\sketchG\) such that \(\contextmorphisma;\contextmorphismr=\contextmorphismt\). Due to Corollary \ref{coro:sketch-morphism-vs-sketch-constraint} and Definition \ref{def:sketch-conditions:satisfaction}, this requirement is obviously equivalent to the statement that \sketchG\ satisfies the global constraint \((uc,\initialmorphism{\contextG})\) with:\\
\(\hspace*{10ex}uc\isdef
\initialobject\contextof 
\forall(\contextL:\guardedexistential{\bigwedge\setofstatementsL}{\contextmorphisma}{\contextR}{\bigwedge\setofstatementsR})\).
\\
Be aware, that the identifier \contextmorphisma\ in  $uc$ does not refer  to the sketch morphism $\contextmorphisma:\sketchL\to\sketchR$ but to the underlying context morphism $\contextmorphisma:\contextL\to\contextR$. Note further, that we can replace \(\setofstatementsR\) by  \((\setofstatementsR\setminus\contextmorphisma(\setofstatementsL))\) without loosing the equivalence!

Further we say  that a sketch \sketchG\ satisfies the "negative universal constraint" $\contextmorphisma:\sketchL\to\sketchR$ if, and only if, for any sketch morphism \(\contextmorphismt:\sketchL\to\sketchG\) there does not exist a sketch morphism \(\contextmorphismr:\sketchR\to\sketchG\) such that \(\contextmorphisma;\contextmorphismr=\contextmorphismt\). This requirement is equivalent  to the statement that \sketchG\ satisfies the global constraint \((nuc,\initialmorphism{\contextG})\) with:\\
\(\hspace*{10ex}nuc\isdef
\initialobject\contextof 
\forall(\contextL:(\bigwedge\setofstatementsL\rightarrow\neg\exists(\contextmorphisma,\contextR:\bigwedge\setofstatementsR)))\)

What can we do if a sketch \sketchG\ does not satisfy a global constraint  \((\condition,\initialmorphism{\contextG})\) for a simple condition of the form \(\condition=\initialobject\contextof 
\forall(\contextL:\guardedexistential{\bigwedge\setofstatementsgen{1}}{\contextmorphisma}{\contextR}{\bigwedge\setofstatementsgen{2}})\) where \setofstatementsgen{1} is a set of statements in \contextL\ and \setofstatementsgen{2} a set of statements in \contextR, respectively? 

We can repair this flaw by applying the corresponding sketch morphism $\contextmorphisma:(\contextL,\setofstatementsgen{1})\to(\contextR,\setofstatementsgen{2}\cup\contextmorphisma(\setofstatementsgen{1}))$ as a transformation rule for all sketch morphisms \(\contextmorphismt:(\contextL,\setofstatementsgen{1})\to\sketchG\) not satisfying the conclusion in condition \condition. In other words, a "match" of the transformation rule is given by a context morphism \(\contextmorphismt:\contextL\to\contextG\) such that \sketchsatisfies{\sketchG}{\bigwedge\setofstatementsgen{1}}{\contextmorphismt} and \sketchsatisfies{\sketchG}{\neg\exists(\contextmorphisma,\contextR:\bigwedge\setofstatementsgen{2})}{\contextmorphismt}. Note, that the "negative application condition" \sketchsatisfies{\sketchG}{\neg\exists(\contextmorphisma,\contextR:\bigwedge\setofstatementsgen{2})}{\contextmorphismt} ensures that we don't apply the rule twice for the same "match" \(\contextmorphismt:(\contextL,\setofstatementsgen{1})\to\sketchG\). Applying the
\\
\begin{minipage}[c]{.4\linewidth}
\vspace*{-1.5ex}
$$\xymatrix{
(\contextL,\setofstatementsgen{1}) \ar[r]^(.36){\contextmorphisma}\ar[d]_(.5){\contextmorphismt}\ar@{}[dr]|{PO}
& (\contextR,\setofstatementsgen{2}\cup\contextmorphisma(\setofstatementsgen{1}))  \ar[d]^(.45){\contextmorphismt^*}
\\
\sketchG\ar[r]^{\contextmorphisma^*} 
& \sketchH
}$$
\end{minipage}
\begin{minipage}[c]{.6\linewidth}
rule $\contextmorphisma$ via the match \contextmorphismt\ means, as usual, nothing but to construct a pushout in the  category \katsketch\ of sketches, as shown on the left. Depending on the property of the context morphism \(\contextmorphisma:\contextL\to\contextR\) the pushout construction may have different effects.  
\end{minipage}
\\
The context \contextG\ can be extended and/or factorized and if \(\setofstatementsgen{2}\not=\emptyset\) we will add new statements to the statements originating from \sketchG. Using constraints we can describe the crucial effect of the rule application as follows: \sketchH\ satisfies the constraint \((\bigwedge\setofstatementsgen{2},\contextmorphismt^*)\) in addition to the constraint \((\bigwedge\setofstatementsgen{1},\contextmorphismt;\contextmorphisma^*)\) inherited from \sketchG.\footnote{It would be good to characterize the class of sketch constraints  preserved by arbitrary sketch morphisms. Probably we should have a closer look at related results in \cite{FreydS1990}. }
\begin{example}[Repairing \(\metasig_{CT}\)-sketches]
\label{ex:repairing-ct-sketches}
As discussed in Example \ref{ex:sketch-constraints-category-theory}, there is one violation of the global constraints \((\conditionct_3,\initialmorphism{\contextG})\) "uniqueness of composition" by the  \(\metasig_{CT}\)-sketch \(\sketchG=(\graphG,\setofstatementsG)\) in Example \ref{ex:dpf-sketches} and one violation of \((\conditionct_6,\initialmorphism{\contextG})\) "decomposition of monomorphisms". Repairing these two violations by pushout construc-

\noindent
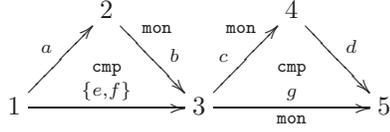
\begin{minipage}[c]{.45\linewidth}
\vspace*{-2ex}
$$\xymatrix{
&  2\ar[dr]^(.6)b\ar@{}[d]|(.6){\featurecomp} \ar@{}[dr]^(.3){\featuremonic} 
&& 4\ar[dr]^d \ar@{}[d]|(.6){\featurecomp}
\\
1\ar[ur]^{a}\ar[rr]^{\{e,f\}}
&{}& 3\ar[ur]^(.4)c\ar@{}[ur]^(.7){\featuremonic} \ar[rr]^g_{\featuremonic}
&{}& 5
}$$
\end{minipage}
\begin{minipage}[c]{.55\linewidth}
tions, as described above, will result in a \(\metasig_{CT}\)-sketch \sketchH\
like the one visualized on the left. \sketchG\ also does not satisfy the global constraint \((\conditionct_2,\initialmorphism{\contextG})\) "definedness of composition" and the global constraint "existence
\end{minipage} 
\\
of identities" that has not been formalized in Example \ref{ex:sketch-constraints-category-theory}. We do not want to require that any \(\metasig_{CT}\)-sketch satisfies these two global constraints since we do not intend to use \(\metasig_{CT}\)-sketches just as encodings of categories but rather as (hopefully finite) representations of (possibly infinte) categories. This is the original purpose of sketches in category theory. See the discussion in Remark \ref{rem:sketch-constraints-in-dpf}.
\qed 
\end{example}	

The utilization of sketch morphisms as "universal constraints" and corresponding transformation rules, as discussed above, allows us, according to Remark \ref{rem:sketch-constraints-colimits}, to express, on one side, arbitrary requirements concerning the existence of  (co)limits and to generate, on the other side, 
\(\metasig_{CT}\)-sketches satisfying a given set of (co)limit requirements. This may be one of the reasons that there is no need for arbitrary first-order sketch conditions in \cite{Mak97}.

\noindent
\textbf{Deduction:}
Generating new statements from given statements by means of rules is the essence of deduction in logic. An interesting observation is that the "repairing procedure", discussed in the last paragraph, can be also described as a procedure deducing new sketch constraints from given sketch constraints.

We consider a sketch \sketchG\ together with a set \setofconstraintsG\ of sketch constraints on \contextG. If \setofconstraintsG\ contains a global constraint \((\condition,\initialmorphism{\contextG})\) with \(\condition=\initialobject\contextof 
\forall(\contextL:\guardedexistential{\bigwedge\setofstatementsgen{1}}{\contextmorphisma}{\contextR}{\bigwedge\setofstatementsgen{2}})\) we can deduce a local constraint \((\guardedexistential{\bigwedge\setofstatementsgen{1}}{\contextmorphisma}{\contextR}{\bigwedge\setofstatementsgen{2}},\contextmorphismt)\) on \contextG\ for any context morphism \(\contextmorphismt:\contextL\to\contextG\). This step corresponds to the \textbf{universal elimination rule} in classical first-order logic. If there is now a constraint \((\bigwedge\setofstatementsgen{1},\contextmorphismt)\in\setofconstraintsG\), we can apply a kind of \textbf{modus ponens rule} and deduce the constraint \((\uncondexists{\contextmorphisma}{\contextR}{\bigwedge\setofstatementsgen{2}},\contextmorphismt)\) on \contextG. 
Keep in mind that \(\contextL\contextof\uncondexists{\contextmorphisma}{\contextR}{\bigwedge\setofstatementsgen{2}}\)!
The pushout construction generates, finally, the constraint \((\bigwedge\setofstatementsgen{2},\contextmorphismt^*:\contextR\to\contextH)\) on \contextH. This looks very much like an analogon to \textbf{Skolemization} in classical first-order logic. More precisely, we can consider this pushout construction as a pendant to the introduction of \textbf{Skolem constants}. This is quite in accordance with the characterization of operations in graph term algebras by pushouts in \cite{WolterDK18}.

As another example motivating the use of sketch constraints as "first class citizens", we discuss  statements in \(\metasig_{CT}\)-sketches, i.e., atomic \(\metasig_{CT}\)-constraints, as introduced and discussed in the Examples \ref{ex:dpf-sketches} and \ref{ex:sketch-constraints-category-theory}: We included the predicate symbols \featuremonic\ and \featurefinal\ in our sample footprint \(\metasig_{CT}\) to exemplify, in a more appropriate way, the use of predicate symbols in diagrammatic specifications in general. In Example \ref{ex:sketch-constraints-category-theory} we discussed, first, that we can specify known or desired properties of predicates by means of sketch conditions. Later, we have shown that we can even express the universal properties defining the concepts "monomorphism" and "final object", respectively, by means of sketch conditions. 

Given a \(\metasig_{CT}\)-sketch \(\sketchG=(\contextG,\setofstatementsG)\), the sketch condition \(\conditionct_7\), defining the concept  monomorphism, may help us to deduce from the \featurecomp-statements, present in \setofstatementsG, that two parallel edges in \contextG\ have to be identified. We need just a rule which generates for each 
atomic \(\metasig_{CT}\)-constraint \((\featuremonic,\bindinga:\arity(\featuremonic)\to\contextG)\) in \setofstatementsG\ a corresponding sketch constraint \((\conditionct_7,\bindinga)\) on \sketchG.\footnote{This works so easy, since we designed our examples in such a way that the context of \(\conditionct_7\) is just \(\arity(\featuremonic)\). In general, any atomic \(\metasig\)-constraint \((\predP,\bindinga:\arity(\predP)\to\contextG)\) and condition \(\contextK\contextof\condition\) may generate a sketch constraint \((\condition,\contextmorphismc;\bindinga)\) for any $\contextmorphismc:\contextK\to\arity(\predP)$. }
Since \bindinga\ binds all "free variables" in \(\conditionct_7\), we need just to adapt the three steps (1) universal elimination, (2) modus ponens  and (3) Skolemization, as discussed above for global constraints, to deduce identifications of parallel edges in \contextG.

To keep \(\metasig_{CT}\) as small as possible, we have not included in  \(\metasig_{CT}\) predicate symbols for other limits and colimits like \featureequ, \featurepb, \featurepo, \featureprod, for example. Utilizing sketch constraints we can even avoid to do this!
In analogy to "anonymous functions" in programming, we can use, according to Remark \ref{rem:sketch-constraints-colimits}, the condition \(C_\graphI\contextof\bigwedge\{exists_\graphI,unique_\graphI\}\) as an \textbf{anonymous predicate} representing the (co)limit concept that corresponds to the  shape graph \graphI. 
With anonymous predicates we can not formulate statements, i.e., entities within a sketch \sketchG, but constraints on the sketch \sketchG.  Note, that we need, of course, conjunction introduction and elimination rules to work properly with sketch constraints of the form \((\bigwedge\{exists_\graphI,unique_\graphI\},\bindinga:C_\graphI\to\contextG)\) or more general \((\bigwedge\setofstatementsgen{},\contextmorphismt)\).

There should be now sufficient evidence that it will be beneficial to work in future DPF with sketch constraints as first class citizens and our discussion suggests, especially, to employ pairs of a sketch $\sketchG=(\contextG,\setofstatementsG)$ and a set \setofconstraintsG\ of sketch constraints on \sketchG\ as an appropriate formalization of software models. We will call those pairs (hierarchical triples) \(((\contextG,\setofstatementsG),\setofconstraintsG)\) \textbf{constrained sketches}.

\begin{remark}[Constrained sketches in MDE]
\label{rem:sketch-constraints-in-dpf}
Our approach to use and develop DPF as a theoretical foundation of MDE is based on the idea that any diagrammatic specification formalism/technique is characterized by a certain choice of a category \katcontext\ and a footprint \metasig\ where the corresponding diagrams/models can be described as \metasig-sketches. 
Sketch conditions and sketch constraints have been developed to provide the necessary additional means   to describe/constrain the syntactic structure of diagrams/models. In such a way, we can characterize now a diagrammatic specification formalism not only by a certain category \katcontext\ and a certain footprint \metasig\ but also by an additional set of \metasig-sketch conditions. 

We should, however, distinguish between two kinds of \metasig-sketch conditions: The first kind of conditions is used to formulate those constraints on \metasig-sketches $\sketchG$ that can be legally used as elements in \setofconstraintsG. 
For a constrained \metasig-sketch \((\sketchG,\setofconstraintsG)\) 
the occurence of a constraint \((\condition,\contextmorphismt)\) in \setofconstraintsG\ will certify that \(\sketchsatisfies{\sketchG}{\condition}{\contextmorphismt}\). Requirements for the "relational data model" \cite{Rutle10,RRLW10_JLAP_NWPT2009} like "every table must have a primary key" and "a foreign key should only refer to a primary key" will be formalized by conditions of this kind. 

Conditions formalizing requirements like "inheritance is transtive" or "a subclass inherits all attributes of all its superclasses", however, should not be included in any \setofconstraintsG\ to avoid that   diagrams/models become too much polluted with redundant information. Those additional conditions are part of the formalism as a whole and represent the background knowledge and rules that can be used to deduce for any constrained sketch information from the  information given in \setofstatementsG\ and \setofconstraintsG, respectively, and to repair violations of the constraints in \setofconstraintsG.
\qed
\end{remark}


\noindent
\textbf{Conceptual hierarchy:}
Introducing "constrained sketches" teleports us "back to start" but on a higher conceptual level.
%
We do have a category \katsketch\ of sketches. To any sketch \(\sketchG=(\contextG,\setofstatementsG)\) we can assign the set \(\functorCstr(\sketchG)\) of all  sketch constraints \((\condition,\contextmorphismt:\contextK\to\contextG)\) on context \contextG\ with \condition\ a first-order sketch condition in \(\functorSC(\contextK)\) according to Definition \ref{def:sketch-conditions:syntax}. Analogously to the translation of atomic \metasig-constraints, described in Example \ref{ex:dpf-sketches}, we can define for any sketch morphism $\contextmorphisma:\sketchG\to\sketchH$ a map \(\functorCstr(\contextmorphisma):\functorCstr(\sketchG)\to\functorCstr(\sketchH)\) by simple post-composition with the underlying context morphism $\contextmorphisma:\contextG\to\contextH$: \(\functorCstr(\contextmorphisma)(\condition,\contextmorphismt)\isdef(\condition,\contextmorphismt;\contextmorphisma)\) for all \((\condition,\contextmorphismt)\in\functorCstr(\sketchG)\). This gives us trivially a functor \(\functorCstr:\katsketch\to\katset\) at hand. 

This situation is, however, just an instance of the "abstract pattern" we started with in Section \ref{sec:abstract-sketches}: The category \katsketch\ can be taken as an instance of \katcontext\ and the functor \(\functorCstr:\katsketch\to\katset\) as an instance of \flar{\functorStm}{\katcontext}{\katset}, respectively. The "constrained sketches" are then nothing but the "abstract sketches" for this instance! We can now consider first-order sketch conditions and sketch constraints for this new instance and will finally get a further instance of the "abstract pattern". Potentially, we can even iterate this procedure ad infinitum. 

Iterating this procedure is maybe not that relevant in practice. We take it, however, as a good sign that our category independent approach allows  us to move in and furnish the next higher level in the conceptual hierarchy whenever it is necessary and/or opportune.

For the moment, we envision three kinds of deduction tasks: (1) Deduction of statements from statements and constraints. Probably the most relevant variant for applications of DPF in MDE. (2) Deduction of constraints from constraints. In category theory we prove, for example, that products and equalizer provide pullbacks. 
(3) Deducing deduction rules from given deduction rules. 

\begin{remark}[Hierarchy of sketches]
\label{rem:conceptual-hierarchy}
How is our "conceptual hierarchy" related to the "hierarchy of sketches" in \cite{Mak97}? Makkai starts with a presheaf topos, i.e., a functor category \(\katcontext=\functorcat{\katC}{\katset}\). Note, that topoi are adhesive! As example we consider 
the presheaf topos
\(\katgraph\cong[\xymatrix{E\ar@(ul,dl)[]_(.5){id_E}\ar@/^/[r]^s\ar@/_/[r]^t&V\ar@(ur,dr)[]^(.5){id_V} }\longrightarrow\katset]\). 

\vspace*{1ex}
Then he describes a nice theoretical result: For any footprint \(\metasig=(\setpred,\arity)\), \(\flar{\arity}{\setpred}{\functorcat{\katC}{\katset}}_{Obj}\) there is a category \(\setpred\overrightarrow{\arity}\katC\) such that the category  \katmsketch\ of multi \metasig-sketches, discussed in Remark \ref{rem:adhesiveness}, is isomorphic to the presheaf topos \functorcat{\setpred\overrightarrow{\arity}\katC}{\katset}.
\footnote{Preparing \cite{DW08} we proved the same result for the general case with \setpred\ a category and \arity\ a functor, but could not include the result and the lengthy proof in  the paper \cite{DW08}.}
\(\setpred\overrightarrow{\arity}\katC\) can be constructed as follows: We take the disjoint union of \setpred (as a discrete category) and \katC. For any predicate symbol \(\predP\in\setpred\), any object \objectC\ in \katC, and any \(\elementc\in\arity(\predP)(\objectC)\) we add an arrow \((\predP,\elementc,\objectC):\predP\to\objectC\). Finally, we define the composition for the new pairs of composable arrows: \((\predP,\elementc,\objectC);\morphismf\isdef(\predP,\arity(\predP)(\morphismf)(\elementc),\objectC')\) for all \(f:\objectC\to\objectC'\) in \katC.
As example, we \nolinebreak take
\\
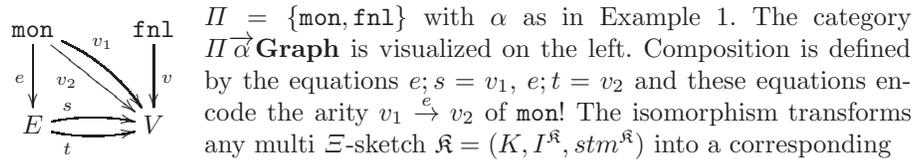
\begin{minipage}[c]{.23\linewidth}
\vspace*{-2ex}
$$\xymatrix{
  \featuremonic\ar[d]_(.5){\edgee}\ar@/^/[dr]^(.4){\vertexv_1}\ar[dr]_(.4){\vertexv_2}
& \featurefinal\ar[d]^{\vertexv} 
\\
  E\ar@/^.3pc/[r]^(.3)s\ar@/_.3pc/[r]_(.3)t 
& V
}$$
\end{minipage}
\begin{minipage}[c]{.77\linewidth}
\(\setpred=\{\featuremonic,\featurefinal\}\) with \arity\ as in Example \ref{ex:dpf-sketches}. The category \(\setpred\overrightarrow{\arity}\katgraph\) is visualized on the left. Composition is defined by the equations \(\edgee;s=\vertexv_1\),  \(\edgee;t=\vertexv_2\) and these equations encode the arity $\vertexv_1\stackrel{\edgee}{\to}\vertexv_2$ of \featuremonic! The isomorphism transforms any multi \metasig-sketch \(\sketchK=(\contextK,I^\sketchK,stm^\sketchK)\) into a corresponding
\end{minipage} 
\\
functor $\calK:\setpred\overrightarrow{\arity}\katgraph\to\katset$. \(\calK(\xymatrix{E\ar@/^.2pc/[r]^s\ar@/_.2pc/[r]_t&V })\) represents the graph \graphK.
The set \(\calK(\featuremonic)\) holds all the identifiers $\elementi\in I^\sketchK$ with \(stm^\sketchK(\elementi)=(\featuremonic,\bindinga)\) while the maps \(\calK(\edgee)\), \(\calK(\vertexv_1)\), \(\calK(\vertexv_2)\) encode all the corresponding bindings \(\bindinga:\arity(\featuremonic)\to\graphK\).

After transforming \katmsketch\ into \functorcat{\setpred\overrightarrow{\arity}\katC}{\katset}, we can define another footprint \(\metasig'=(\setpred',\arity')\), \(\flar{\arity'}{\setpred'}{\functorcat{\setpred\overrightarrow{\arity}\katC}{\katset}}_{Obj}\) on this next level of the hierarchy and start again but this time with atomic \(\metasig'\)-constraints as statements.

There are no sketch conditions in \cite{Mak97} but any multi \metasig-sketch \(\sketchK=(\contextK,I^\sketchK,stm^\sketchK)\) correponds to the sketch condition \(\contextK\contextof\bigwedge\{stm^\sketchK(\elementi)\mid\elementi\in I^\sketchK\}\). In such a way, all the arities \(\arity'(\predP')\) in the footprint \(\metasig'\) correspond to very simple sketch conditions that are just conjunctions of \metasig-statements and atomic \(\metasig'\)-constraints correspond to sketch constraints employing only those conditions, of this simple kind, that correspond to arities \(\arity'(\predP')\) and each condition gets the "label" \predP'. We propose to work, instead, with arbitrary un-labeled first-order sketch conditions.

Apropos  "nice theoretical result": For the footprint \(\metasig_R=(\mathsf{Rel},\arity)\) in Ex.~\ref{ex:gratra-sketches} we can consider \arity\ as a map \(\arity:\mathsf{Rel}\to[\katgen{1}\rightarrow\katset]_{Obj}\) with  $V$ the only object in \katgen{1} and \(\arity(\predP)(V)=\{x_1,x_2\}\) for all \(\predP\in\mathsf{Rel}\). \(\setpred\overrightarrow{\arity}\katgen{1}\)  contains then for each \(\predP\in\mathsf{Rel}\)
\\
\begin{minipage}[c]{.23\linewidth}
\vspace*{-1.5ex}
$$\xymatrix{
\cdots\;\; \predP\ar@/^.3pc/[r]^(.5){(\predP,x_1)}\ar@/_.3pc/[r]_(.5){(\predP,x_2)}
& V
}$$
\end{minipage}
\begin{minipage}[c]{.77\linewidth}
an "edge sort" \predP\ and  \functorcat{\setpred\overrightarrow{\arity}\katgen{1}}{\katset} is the category of graphs with an \(\mathsf{Rel}\)-indexed family of edges. This category is adhesive in contrast to the category of \(\mathsf{Rel}\)-labelled graphs in \cite{Rensink04}.
\qed
\end{minipage} 
\end{remark}

\section{Conclusions and future work}
\label{sec:conclusions}
We presented a universal method to define a category \katsketch\ of abstract sketches (specifications, models) for arbitrary categories \katcontext\ and "statement" functors \flar{\functorStm}{\katcontext}{\katset} as well as corresponding general first-order conditions and constraints on those sketches. We verified that our method is indeed a generalization of different variants of "graph conditions and constraints" \cite{BrugginkCHK11,EEPT06,HabelP09,Kosiol0TZ20,Rensink04} and of "universal constraints" and "negative universal constraints" in DPF \cite{Rutle10,RRLW10_JLAP_NWPT2009}.

We exemplified the usefulness of general first-order constraints to describe the syntactic structure of sketches as well as to deduce information from the information given in a sketch. We discussed and exemplified vital observations, insights and ideas concerning a future deduction calculus for sketch constraints. Sketch constraints are statements about sketches thus our universal method can be applied iteratively. 
The category \katsketch\ can be taken as an instance of \katcontext\ and the functor \(\functorCstr:\katsketch\to\katset\), assigning to each sketch \sketchG\ the set \(\functorCstr(\sketchG)\) of all  sketch constraints on \sketchG, as an instance of \flar{\functorStm}{\katcontext}{\katset}, respectively. This gives us open conceptual hierarchies of sketches for a future further development of DPF
at hand.

We mention only four of the "work packages" waiting for us: (1) Developing a universal deduction calculus for sketch constraints, (2) Revising and enhancing the meta-modeling facilities of DPF, (3) Revising and further developing \cite{Wolter2021logics,WolterDK18} (4) Integrating everything. Besides this it may be worth to think about diagrammatic versions of OCL and Description Logics, for example, and to develop tools for diagrammatic reasoning, especially for teaching Category Theory. 


\bibliographystyle{splncs04}
\bibliography{icgt-2021}

\begin{thebibliography}{10}
\providecommand{\url}[1]{\texttt{#1}}
\providecommand{\urlprefix}{URL }
\providecommand{\doi}[1]{https://doi.org/#1}

\bibitem{BrugginkCHK11}
Bruggink, H.J.S., Cauderlier, R., H{\"{u}}lsbusch, M., K{\"{o}}nig, B.:
  {Conditional Reactive Systems}. In: Chakraborty, S., Kumar, A. (eds.)
  {FSTTCS} 2011 Proceedings, Mumbai India. LIPIcs, vol.~13, pp. 191--203.
  Schloss Dagstuhl - Leibniz-Zentrum f{\"{u}}r Informatik (2011).
  \doi{10.4230/LIPIcs.FSTTCS.2011.191}

\bibitem{DW08}
Diskin, Z., Wolter, U.: {A Diagrammatic Logic for Object-Oriented Visual
  Modeling}. ENTCS  \textbf{203/6},  19--41 (2008).
  \doi{10.1016/j.entcs.2008.10.041}

\bibitem{EEPT06}
Ehrig, H., Ehrig, K., Prange, U., Taentzer, G.: {Fundamentals of Algebraic
  Graph Transformations}. EATCS Monographs on Theoretical Computer Science,
  Springer, Berlin (2006). \doi{10.1007/3-540-31188-2}

\bibitem{FreydS1990}
Freyd, P.J., Scedrov, A.: Categories, allegories, North-Holland Mathematical
  Library, vol.~39. North-Holland (1990)

\bibitem{HabelP09}
Habel, A., Pennemann, K.: Correctness of high-level transformation systems
  relative to nested conditions. Math. Struct. Comput. Sci.  \textbf{19}(2),
  245--296 (2009). \doi{10.1017/S0960129508007202}

\bibitem{Kosiol0TZ20}
Kosiol, J., Str{\"{u}}ber, D., Taentzer, G., Zschaler, S.: {Graph Consistency
  as a Graduated Property - Consistency-Sustaining and - Improving Graph
  Transformations}. In: Gadducci, F., Kehrer, T. (eds.) {ICGT} 2020,
  Proceedings. LNCS, vol. 12150, pp. 239--256. Springer (2020).
  \doi{10.1007/978-3-030-51372-6\_14}

\bibitem{Mak97}
Makkai, M.: Generalized sketches as a framework for completeness theorems.
  Journal of Pure and Applied Algebra  \textbf{115},  49--79, 179--212,
  214--274 (1997)

\bibitem{Rensink04}
Rensink, A.: {Representing First-Order Logic Using Graphs}. In: Ehrig, H.,
  Engels, G., Parisi{-}Presicce, F., Rozenberg, G. (eds.) {ICGT} 2004. LNCS,
  vol.~3256, pp. 319--335. Springer (2004). \doi{10.1007/978-3-540-30203-2\_23}

\bibitem{RLGRW14_JFAC}
Rossini, A., de~Lara, J., Guerra, E., Rutle, A., Wolter, U.: A formalisation of
  deep metamodelling. Formal Aspects of Computing pp. 1--38 (2014).
  \doi{10.1007/s00165-014-0307-x}

\bibitem{RRLW09_JLAP}
Rossini, A., Rutle, A., Lamo, Y., Wolter, U.: {A formalisation of the
  copy-modify-merge approach to version control in MDE}. Journal of Logic and
  Algebraic Programming  \textbf{79}(7),  636--658 (2010).
  \doi{10.1016/j.jlap.2009.10.003}

\bibitem{Rutle10}
Rutle, A.: {Diagram Predicate Framework: A Formal Approach to MDE}. Ph.D.
  thesis, Department of Informatics, University of Bergen, Norway (2010),
  \url{https://hdl.handle.net/1956/4469}

\bibitem{RRLW10_JLAP_NWPT2009}
Rutle, A., Rossini, A., Lamo, Y., Wolter, U.: {A formal approach to the
  specification and transformation of constraints in MDE}. Journal of Logic and
  Algebraic Programming  \textbf{81/4},  422--457 (2012).
  \doi{10.1016/j.jlap.2012.03.006}

\bibitem{WolterMH2020}
Wolter, U., Martini, A.R., Haeusler, E.H.: {Indexed and Fibred Structures for
  Hoare Logic}. Electronic Notes in Theoretical Computer Science (348),
  125--145 (2020). \doi{10.1016/j.entcs.2020.02.008}

\bibitem{Wolter2021logics}
Wolter, U.: {Logics of First-Order Constraints - A Category Independent
  Approach} (Jan 2021), \url{https://arxiv.org/abs/2101.01944}

\bibitem{WolterDK18}
Wolter, U., Diskin, Z., K{\"{o}}nig, H.: {Graph Operations and Free Graph
  Algebras}. In: Graph Transformation, Specifications, and Nets. pp. 313--331.
  Springer, LNCS 10800 (2018). \doi{10.1007/978-3-319-75396-6\_17}

\bibitem{WolterM2013}
Wolter, U., Mantz, F.: {The Diagram Predicate Framework in View of Adhesive
  Categories}. Tech. Rep.~358 (Aug 2013),
  \url{http://www.ii.uib.no/publikasjoner/texrap/pdf/2013-405.pdf}

\end{thebibliography}
\end{document}